\documentclass[10pt]{article}  

\usepackage{amsmath}
\usepackage{amssymb}

\usepackage{graphicx}

\usepackage{cite}

\usepackage{color} 

\usepackage[labelfont=bf,labelsep=period,justification=raggedright]{caption}

\makeatletter
\renewcommand{\@biblabel}[1]{\quad#1.}
\makeatother

\date{}

\begin{document}

\begin{flushleft}
{\Large
\textbf{Statistical Signs of Social Influence on Suicides}
}
\\
Hygor Piaget M. Melo$^{1\ast}$, 
Andr\'e A. Moreira$^{1}$,
Hern\'an A. Makse$^{2}$,
Jos\'e S. Andrade Jr.$^{1}$
\\

\bf{1} Departamento de F{\'i}sica, Universidade Federal do Cear{\'a}, 60451-970 Fortaleza, Cear{\'a}, Brazil 
\\
\bf{2} Levich Institute and Physics Department, City College of New York, New York, NY 10031, USA
\\

$\ast$ E-mail: Corresponding hygor@fisica.ufc.br
\end{flushleft}

\section*{Abstract}
Certain currents in sociology consider society as being composed of
autonomous individuals with independent psychologies. Others, however,
deem our actions as strongly influenced by the accepted standards of
social behavior. The later view was central to the positivist
conception of society when in 1887 \'Emile Durkheim published his
monograph {\it Suicide} \cite{Durkheim1897}. By treating the suicide as a social fact,
Durkheim envisaged that suicide rates should be determined by the
connections (or the lack of them) between people and society. Under
the same framework, Durkheim considered that crime is bound up with
the fundamental conditions of all social life and serves a social
function. In this sense, and regardless of its extremely deviant
nature, crime events are somehow capable to release certain social
tensions and so have a purging effect in society. The social effect on
the occurrence of homicides has been previously substantiated
\cite{Bettencourt2007,Alves2013}, and confirmed here, in terms of a superlinear
scaling relation: by doubling the population of a Brazilian city
results in an average increment of $135\;\%$ in the number of
homicides, rather than the expected isometric increase of $100\;\%$,
as found, for example, for the mortality due to car crashes. Here we
present statistical signs of the social influence on the suicide
occurrence in cities. Differently from homicides (superlinear) and
fatal events in car crashes (isometric), we find sublinear scaling
behavior between the number of suicides and city population, with
allometric power-law exponents, $\beta=0.836\pm 0.009$ and $0.870\pm
0.002$, for all cities in Brazil and US, respectively. The fact that
the frequency of suicides is disproportionately small for larger
cities reveals a surprisingly beneficial aspect of living and
interacting in larger and more complex social networks.

\section*{Author Summary}

\section*{Introduction}

It is not uncommon in nature to observe properties that present
non-trivial forms of scale dependence. This is the case, for instance,
of critical phenomena, where scaling invariance, universal properties
and renormalization concepts constitute the theoretical framework of a
well-established field in physics \cite{Stanley87}. In biology, the
so-called allometric relations certainly represent outstanding
examples of natural scaling laws. Precisely, allometry implies the use
of power-laws, $Y\propto M^\beta$, to describe the dependence of a
wide range of anatomical, physiological and behavioral properties,
denoted here as $Y$, on the size or the body mass of different animal
species, $M$. If the scaling exponent is $\beta=1$, the
variables $Y$ and $M$ are trivially proportional, and the relation
between them is said to be isometric, while $\beta \neq 1$ indicates
an allometric type of relationship. 
 The so-called ``three-quarters law'' or
Kleiber's law, as originally proposed by the agricultural biologist
Max Kleiber in 1947 \cite{Kleiber1947}, is surely one of the most prominent
allometric relations found in natural sciences. Based on an extensive
set of experimental data, this fundamental law states that the
metabolic rate of all animals should scale to the $3/4$ power of their
corresponding masses \cite{Kleiber1961,Schimidt1984}.

In analogy with biological scaling laws, Bettencourt \emph{et al.}
\cite{Bettencourt2007} showed that, regardless the enormous complexity
and diversity of human behavior and the extraordinary geographic
variability of urban settlements, cities belonging to the same urban
system obey pervasive allometric relations with population size,
therefore exhibiting nonextensive rates of innovation, wealth
creation, patterns of consumption, human social behavior, and several
other properties related to the urban infrastructure.  The authors
then conclude that all data can be grouped into three categories,
namely, material infrastructure, individual human needs, and patterns
of social activity. Despite the unambiguous presence of power-laws,
the urban indicators do not necessarily follow a universal
behavior. Instead the results can be divided in three different
classes. The isometric (linear) case ($\beta=1$) typically reflects
the scaling of individual human needs, like the number of jobs,
houses, and water consumption. As in biology, the allometric sublinear
behavior ($\beta<1$) implies an economy of scale in the quantity of
interest, because its {\it per capita} measurement decreases with
population size. In the case of cities, this is materialized, for
example, in the number of gasoline stations, the total length of
electrical cables, and the road surface (material and
infrastructure). The case of superlinear allometry ($\beta>1$) in
urban indicators emerges whenever the complex patterns of social
activity have significant influence. Wages, income, growth domestic
product, bank deposits, as well as rates of invention, measured by new
patents and employment in creative sectors, all display a superlinear
increase with population size \cite{Bettencourt2007}. While these
results indicate that larger cities are associated with higher levels
of human productivity and quality of life, superlinear scaling can
also characterize negative urban scenarios, such as the prospect of
living costs, crime rates, pollution and disease incidence
\cite{Bettencourt2007,Bettencourt2010a,Bettencourt2010b,Alves2013}.

\section{Materials and Methods}

\noindent\textbf{Brazilian Data.}
 We analyzed data available for all Brazilian cities from 1992 to 2009, 
	made freely available by the Brazil's public healthcare system –
DATASUS \cite{DATASUS}. Here we consider that cities are the smallest administrative
units with local government. The data consist of the number of homicides, suicides, and
deaths in traffic accidents as well as the population for each city.

\noindent\textbf{US Data.}
We used data from the National Cancer Institute SEER, Surveillance Epidemiology and 
End Results downloaded from http://seer.cancer.gov/data/.
The Institute provides mortality data aggregated for three or five years. 
To compare with Brazil, we use suicide mortality data for each American county 
accumulated through the five years of 2003 to 2007. Since the population 
is almost constant for a five years period, we adopted the average county 
population as a measurement for the population on the allometry relation.

\begin{figure*}[!ht]
\includegraphics*[width=16.0cm]{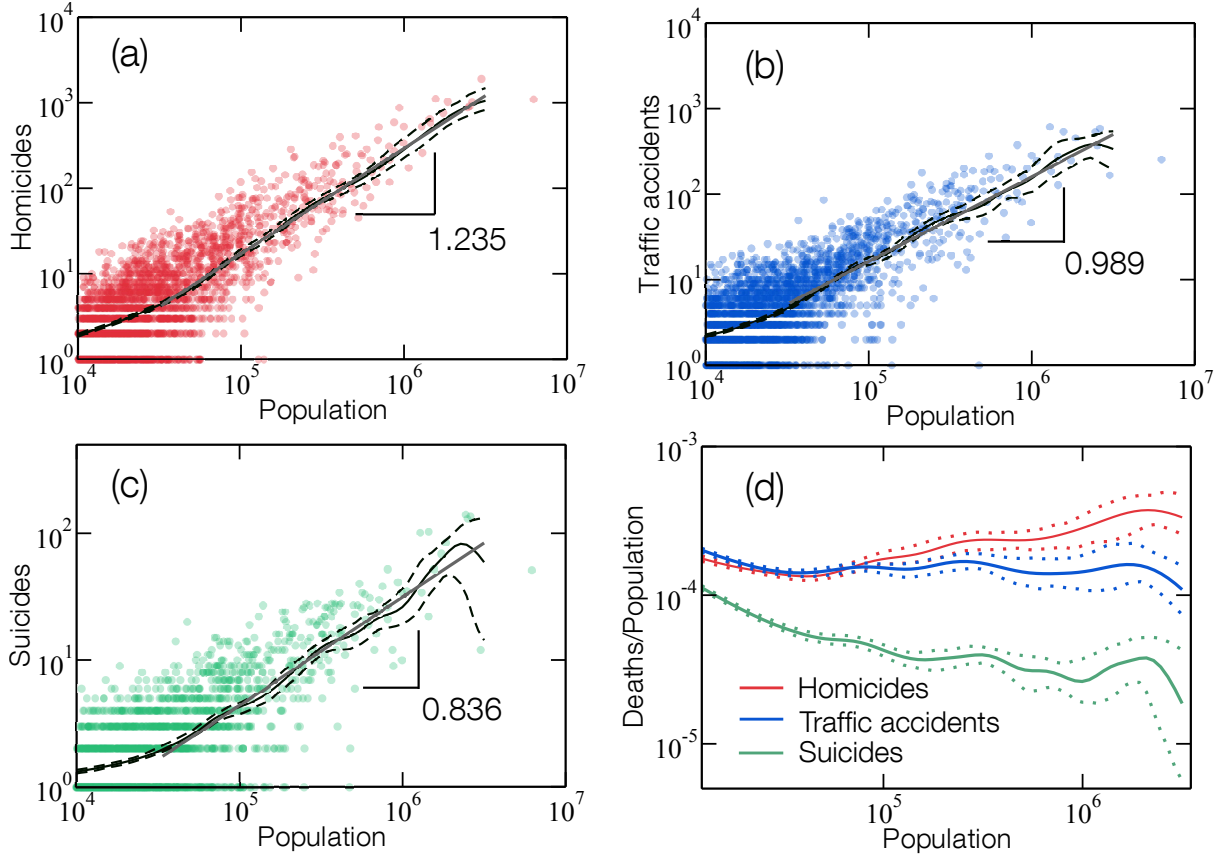}
\label{fig:scaling}
\caption{ \textbf{Scaling relations for homicides, traffic
    accidents, and suicides for the year of 2009 in Brazil.}  The
  small circles show the total number of deaths by (a) homicides (red),
  (b) traffic accidents (blue), and (c) suicides (green) vs the population of each city.
  Each graph represents only one urban indicator, and
  the solid gray line indicate the best fit for a power-law relation, using OLS regression, between the total number of deaths and the city size
  (population). 
  To reduce the fluctuations we also performed a
  Nadaraya-Watson kernel regression \cite{Nadaraya1964,Watson1964}. The dashed lines show the 95\% 
	confidence band for the Nadaraya-Watson kernel regression.
The ordinary least-squares (OLS) \cite{Montogomery1992} fit to the data of homicides in (a) 
reveals an allometric exponent $\beta=1.235 \pm 0.005$, with a 95\% 
confidence interval estimated by bootstrap. This is compatible with 
previous results obtained for U.S. \cite{Bettencourt2007} that also indicate a super-linear
 scaling relation with population and an exponent $\beta=1.16$. Using the 
same procedure, we find $\beta=0.989 \pm 0.007$ and $0.836 \pm 0.009$ for the 
numbers of deaths in traffic accidents (b) and suicides (c), respectively. 
This non-linear behavior observed for homicides and suicides certainly reflects 
the complexity of human social relations and strongly suggests that the the 
topology of the social network plays an important role on the rate of these 
events. (d) The solid lines show the Nadaraya-Watson kernel regression rate 
of deaths (total number of deaths divided by the population of a city) for each
 urban indicator, namely, homicides (red), traffic accidents (blue), and suicides (green).
 The dashed lines represent the 95\% confidence bands. While the rate of fatal 
traffic accidents remains approximately invariant, the rate of homicides systematically 
increases, and the rate of suicides decreases with population.}

\end{figure*}

\section{Results}

\noindent\textbf{Allometry in Urban Indicators.}
The main goal here is to investigate the scaling behavior with city
population of three urban indicators, namely the number of homicides,
deaths in traffic accidents and suicides. For this, we analyzed data
available for all Brazilian cities and 
as well as suicide for US counties, as presented 
previously on Materials and Methods. For 2009 in Brazil, as shown in Fig.~1, the
increase in the number of casualties $D$ with city population $P$ for
the three death causes can be properly described in terms of
power-laws, $D=D_{0} P^{\beta}$, where $D_{0}$ is a normalization
constant, and the exponent $\beta$ reflects a global property at play
across the urban system. Interestingly, while the number of deaths by
traffic accidents display isometric scaling, $\beta \approx 1$,
homicides and suicides are both allometric, but obeying superlinear
and sublinear scaling with population, respectively. These results
suggest that the decision to commit a crime or to suicide, instead
 of being purely a consequence of individual choices, might have strong
correlations with the underlying complex social organization and
interactions. This does not seem to be the case of traffic accidents,
since the strong evidence for isometric scaling, $\beta=0.989 \pm 0.007$,
indicates that such events should result from random processes, i.e., 
no social relations need to be implied among the involved people.

\begin{figure*}[!ht]
\includegraphics*[width=7.0cm]{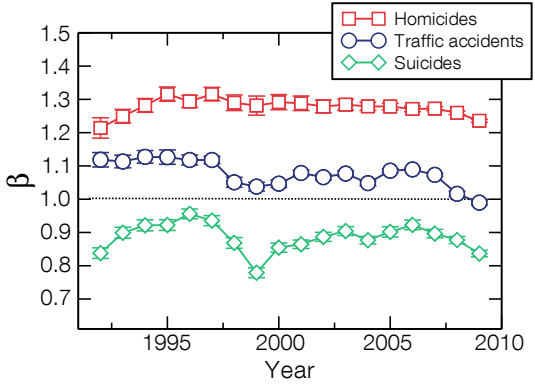}
\label{fig:expoente por ano}
\caption{ \textbf{Temporal evolution of allometric exponent $\beta$ for homicides (red squares),
	deaths in traffic accidents (blue circles), and suicides (green diamonds).} Time evolution
  of the power-law exponent $\beta$ for each behavioral urban
  indicator in Brazil from 1992 to 2009. We can see that the
  non-linear behavior for homicides and suicides are robust for this
  19 years period, and for the traffic accidents the exponent remain
  close of $1.0$.}
%
%
\end{figure*}

From the superlinear scaling exponent found for the number of
homicides in Brazil at the year 2009, $\beta=1.235 \pm 0.005$, one should
expect that, by doubling the population of a city, the number of
homicides would grow approximately by a factor of $135\%$ in average,
instead of just growing $100\%$, as if we had an isometric scaling.
This super-linear behavior is consistent with that found for serious crimes
in USA \cite{Bettencourt2007}. The result obtained for suicide scaling
seems quite surprising. As depicted in
Fig.~1c, the number of suicides scales with the
population size as a power law with exponent $\beta=0.836\pm 0.009$,
which implies an ``economy of scale'' of $22\%$ in comparison with
isometric scaling, similar to Kleiber's law for metabolic rates and
animal masses. This sublinear 
behaviour is reminiscent of
the seminal study by {\'E}mile Durheim \cite{Durkheim1897}, one of the
fathers of modern sociology. In his book \emph{Suicide}, Durkheim
explored the differences in average suicide rates among Protestants
and Catholics, arguing that stronger social control among Catholics
leads to lower suicide rates. The crucial contribution from Durkheim
was certainly to treat the suicide as a social fact, by explaining
variations in its rate at a macro level as a direct consequence of
society-scale phenomena, such as lack of connections between people    
(group attachment) and lack of regulations of behavior, rather than
individuals' feelings and motivations.

The discrepancies observed in the scaling behaviors of homicides,
deaths in traffic accidents and suicides become even more evident if
we plot the average number of deaths per capita against city
population, as shown in Fig.~1d.
Under this framework, the systematic decrease in suicide rate with
population indicates that a large supply of potential social contacts
and interactions might work as an ``antidote'' for this tragic event.
This result is consistent with the idea that human happiness is more a
collective phenomenon than a consequence of individual well-being
conditions. In analogy with health, it is then possible to consider 
a ``happiness epidemy'' spreading in a social network like showed 
previously in \cite{Fowler2008}.
In Fig.~2 we show the dependence on
time of the exponent $\beta$ for a period of 18 years, from 1992 to
2009. We see a robust behavior for $\beta$, in such a way that even
for different years we still having $\beta> 1.0$ for homicides,
$\beta<1.0$ for suicides, and $\beta$ slightly above $1.0$ for deaths by
traffic accidents.

We also analyzed data available for suicides for all US counties. The data 
is an accumulation of the total number of suicides during a period of five years,
from 2003 to 2007. In Fig.~3 we see the total number of suicides
during these five years against the average population for each county.
As we can see the number of suicides also scales with a sub-linear power law with an 
exponent $\beta=0.870 \pm 0.002$. This results corroborates the previously found to Brazil,
indicating that suicide need to be treat as a social fact.

In Fig.~4 we show the density map in the 2D plane,
fatality per capita (deaths divided by population) versus population size for Brazil in 2009. To obtain the approximate
density we perform kernel density estimation in the log-log space. We
also include in the figure lines indicating the 10\%, 50\%, and 90\%
levels for each population size.  Besides confirming the superlinear,
linear, and sublinear behaviors, this results also show how the
probability distribution of rates of fatality vary with the population
size. Also, the 10\% and 90\% lines are representative of expected
extreme cases of low and high fatalities, respectively.

\begin{figure*}[!ht]
\includegraphics*[width=7.0cm]{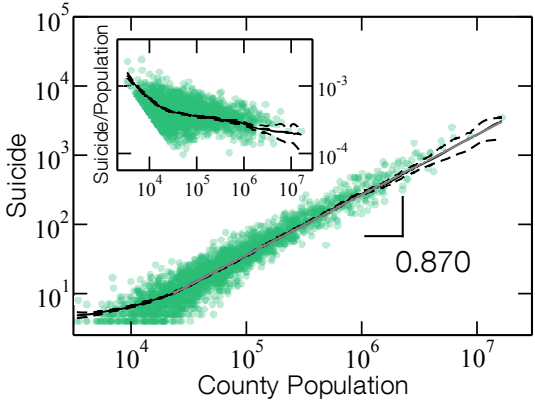}
\label{fig:suicideUS}
\caption{\textbf{Scaling relationship between suicides and population for US counties.} 
	The small circles show the total number of suicides over five years 
	(2003 to 2007) vs the average population of the county. 
	The solid red line indicate the best fit to the data of a power law,
	using OLS regression, between the total number of suicides and the county population. 
    The dashed lines (blue) delimit the 95\% 
	confidence band given by the Nadaraya-Watson kernel regression \cite{Nadaraya1964,Watson1964}.
	By an ordinary least-squares (OLS) \cite{Montogomery1992} fit we find an allometric exponent
  $\beta=0.870 \pm 0.002$ with a $95\%$ confidence interval estimated by bootstrap.
	The inset shows the dependence on population of the suicide rate of the counties
	with the corresponding non-parametric Nadaraya-Watson regression. From this plot, 
	we can clearly observe the systematic decrease of suicide rate with county population.
}
\end{figure*}

\begin{figure*}[!ht]
\includegraphics*[width=17.0cm]{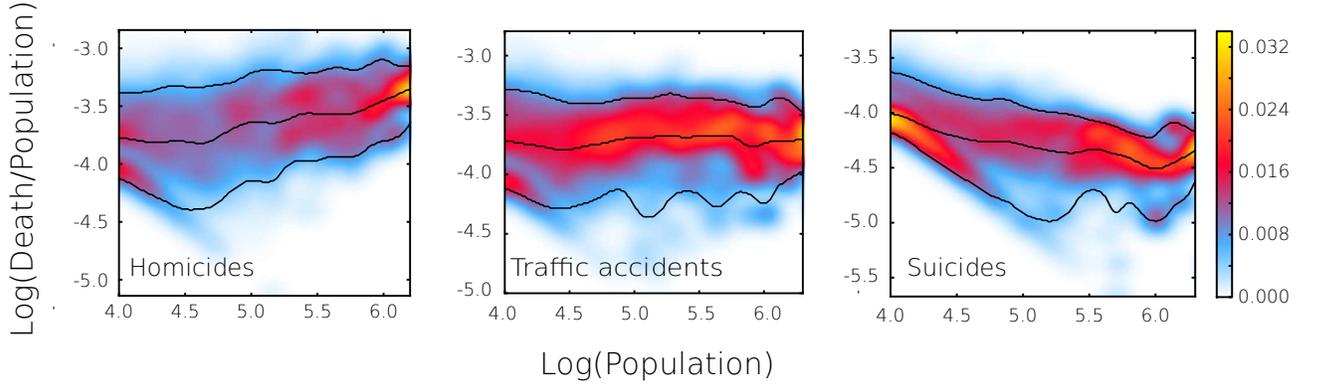}
\label{fig:density}
\caption{ \textbf{Fatality per capita versus population for homicides,
  traffic accidents, and suicides.}
  The color map represents the conditional probability density 
  obtained by kernel density estimation. The bottom and top lines 
  correspond to the  10\% and 90\% bounds of the distribution for each
  population size, that is 80\% of the sampled points are between
  these lines. The middle line is the 50\% level or "median" expected for each
  population size.  The diagonal shape observed in the left side of
  density maps are cases of low number of fatal events, one or two
  fatalities. After this region we observe that the three level lines
  wiggle around an average power-law behavior.  In the case of
  homicides the three level lines indicate an increase in the expected
  density of fatality with the population size.  Similarly, for traffic
  accidents the lines are close to horizontal, that is, the
  probability distribution for the rate of fatality is near
  independent of the population. For suicides, the median
  show a slight decrease with population size, while the 90\% level,
  that is associated with cases of extreme rates of suicide, show a pronounced
  decrease. The sublinear growth observed for suicides, as depicted in Fig.~1c, is likely due
  to the suppression of these extremely high rates in large urban areas.}
\end{figure*}

\section{Discussion}
Allometric relations are ubiquitous in Nature, appearing in a wide variety of biological, sociological, 
chemical and physical systems \cite{West1997,Bettencourt2007,Stanley87}. Even the arrangement of Lego pieces has been recently reported 
to obey a sublinear scaling between the number of pieces types vs. the total number of pieces for many Lego sets \cite{Changizi2002}. 
As originally proposed in biology by Kleiber \cite{Kleiber1947} and extended later by others \cite{West1997}, a single power law 
comprising 27 orders of magnitude can associate the metabolisms of a microscopical bacteria and a blue 
whale, weighting a few picograms and more than 100 t, respectively. Unlike allometric relations 
in Biology, few attempts have been made to explain the origin of universal scaling laws describing
urban indicators. In a recent study ~\cite{Bettencourt2013}, a quantitative framework has been
developed to consider the interactions between people over a social network that is capable to 
predict the allometric scaling for urban systems.

Here we studied the scaling relation with population of three behavioral urban indicators, namely, 
number of homicides, victims of vehicles crashes, and suicide events. We show that, unlike the incidence 
of vehicles crashes with fatal victims,  which exhibits isometric (linear) scaling, homicides and suicides are 
characterized by allometric behaviors, sublinear and superlinear, respectively.

Precisely, we found that the superlinear scaling relation for the 
number of homicides has an allometric exponent that lies between 
1.21 and 1.31 along the years, between 1992 and 2009. Despite all their positive 
attributes \cite{Bettencourt2010a,Bettencourt2010b} (e.g., higher incomes and levels of creative activities),  
these results show that larger and urbanized cities also have a dark side, 
namely, higher levels of violence. Interestingly, the effect is exactly the 
opposite in the number of suicide events, which typically follow a sublinear 
scaling with population, with an allometric exponent value that varies in 
the range between 0.78 and 0.95.
Such a result led us to the conclusion that a suicide event should not 
be taken as an isolated individual decision. This is consistent 
with the conceptual framework put forward by Durkheim \cite{Durkheim1897}, under 
which suicides need to be treated as social facts, actually affected 
by complex human relations. The sublinear behavior found can be 
even considered as a counter-intuitive result. The more common and 
straightforward view would be to associate the suicide causes uniquely 
to a health condition of psychological or mental illness that could 
nevertheless be strongly linked to external urban factors. For instance,
traffic jams, pollution, and stressful jobs all create a harmful environment 
for the population. However, the fact that we found sublinearity, namely, 
that the number of suicides is disproportionately small for larger cities, 
discloses an entirely different perspective. We conjecture that this 
property can be intimately related with an ``emotional epidemy'' as previously 
hypothesized in Ref.~\cite{Fowler2008,Hill2010}. This phenomenon can explain the systematic  
attenuation through the social network of contagious emotions 
states of potentially suicidal individuals.

\section*{Acknowledgments}
We thank the Brazilian Agencies CNPq, CAPES, FUNCAP and FINEP, the
  FUNCAP/CNPq Pronex grant, and the National Institute of Science and
  Technology for Complex Systems in Brazil for financial support.


\end{document}